\documentclass[12pt]{jpconf}

\usepackage{graphicx}%
\usepackage{dcolumn}%
\usepackage{bm}%

\usepackage{subfigure}  

\usepackage{enumerate}

\usepackage{amsgen}
\usepackage{amsfonts}
\usepackage{amsbsy}
\usepackage{amssymb}

\usepackage{amsmath}
\usepackage{amsthm}

\usepackage{braket}
\usepackage{soul}
\setul{0.4pt}{.4pt}

\begin{document}

\title{Faddeev calculations on lambda hypertriton with potentials from Gel'fand-Levitan-Marchenko theory}
\author{E F Meoto and M L Lekala}
\address{Department of Physics, University of South Africa, \\ Private Bag X6, 1710 Johannesburg, South Africa}
\eads{\mailto{EmileMeoto@aims.ac.za}}

\begin{abstract}
Effective lambda-proton and lambda-neutron potentials, restored from theoretical scattering phases through Gel'fand-Levitan-Marchenko theory, are tested on a lambda hypertriton through three-body calculations. The lambda hypertriton is treated as a three-body system consisting of lambda-proton, lambda-neutron and proton-neutron subsystems. Binding energy and root-mean-square radius are computed for the ground state of lambda hypertriton ($J^{\pi}=1/2^+$).  In coordinate space, the dynamics of the system is described using a set of coupled hyperradial equations obtained from the Differential Faddeev Equations. By solving the eigenvalue problem derived from this set of coupled hyperradial equations, the binding energy and root-mean-square matter radius computed are found to be -2.462 MeV and 7.00 fm, respectively. The potentials are also shown to display a satisfactory convergence behaviour.
\end{abstract}

\section{Introduction}
Lambda hypertriton and other light hypernuclei play an important role as femtoscale laboratories for testing the accuracy of new hyperon-nucleon and hyperon-hyperon potentials. In helium bubble-chamber experiments and emulsion experiments carried out over the years since 1952, a large number of lambda hypernuclei have been observed \cite{dav2005, dal2005}. As a result of this large number of lambda hypernuclei, the lambda-nucleon interaction has received considerable attention over the last half-century. 

The lambda-nucleon potentials in active use have their origin in Meson-Exchange SU(3) theory \cite{deSwart1971, deSwart1996, rij1993,rij2001, hol1989,reu1992,hai2005}, Meson-Exchange SU(6) theory \cite{fuj1996a, fuj1996b, fuj2001} or Chiral Effective Field theory \cite{pol2006, pol2007, hai2013}. In order to test their accuracy, these potentials have been used in calculations to compute some important structural properties of light hypernuclei. Some of these properties include the binding energy, lifetime, and root-mean-square radius of lambda hypertriton ($_{\Lambda}^3\text{H}$). The importance of lambda hypertriton in the testing of lambda-nucleon potentials is similar to that of triton and deuteron for nucleon-nucleon potentials. Charge Symmetry Breaking, which is very significant in the lambda-nucleon force, is usually tested by computing the lambda separation energy of the isospin doublet helium-4-lambda ($_{\Lambda}^4\text{He}$) and hydrogen-4-lambda ($_{\Lambda}^4\text{H}$). A range of discrepancies, some negligible and others significant, are observed between these computations and experimental observations. For example, lambda hypertriton lifetimes observed in experiments are about 30 - 50\% shorter \cite{ada2018} than values computed with some lambda-nucleon forces \cite{gal2018b}. In fact, the lifetime of lambda hypertriton is still a puzzle, as there are disagreements even between various experimental results \cite{tro2018,xu2017}.

The significant differences observed between theoretical predictions and experiments suggests that existing theories for the lambda-nucleon force require improvements. Data from more accurate experiments are indispensable to achieving these improvements. These differences also suggest that new perspectives from alternative theories may be needed to \textit{complement} existing theories. In line with the quest for alternative theories for the lambda-nucleon force, new lambda-proton and lambda-neutron potentials were developed in \cite{meo2019}, through the application of Gel'fand-Levitan-Marchenko theory. The aim of this paper is to test the accuracy of the potentials developed in \cite{meo2019}, by computing the binding energy and root-mean-square radius of a lambda hypertriton. The lambda hypertriton is treated as a proton+neutron+lambda three-body system, and the computations are done using the Differential Faddeev Equations in hyperspherical variables.

\section{Hyperspherical harmonics method: Coupled hyperradial equations}

The method used in this paper is fully described in \cite{tho2004}. Therefore, only a brief outline is presented here. Consider a three-body system with particles of masses $m_1, m_2$ and $m_3$ having position vectors $\vec{r}_1, \vec{r}_2$ and $\vec{r}_3 \in \mathbb{R}^3$, respectively. The masses of the particles are in atomic mass units (u). Furthermore, let reduced masses be defined as $A_1=m_1/m, A_2=m_2/m$ and $A_3=m_3/m$, where $m$ is a unit mass, taken here to be the mass of a nucleon. Mass-scaled Jacobi coordinates ($\vec{x}_i$, $\vec{y}_i$), after the elimination of centre-of-mass motion, are defined as follows \cite{tho2004}:
\begin{align}
\label{eq:jacobi_coord}
\vec{x}_i &= \sqrt{\frac{A_j A_k}{A_j + A_k}} \left(\vec{r_j} -  \vec{r_k} \right), \\
\vec{y}_i &= \sqrt{\frac{A_i (A_j + A_k)}{A_i + A_j+ A_k}} \left( \vec{r_i} - \frac{A_j \vec{r_j} + A_k \vec{r_k}}{A_j + A_k} \right), 
\end{align}
where $i,j,k \in (1,2,3)$. In the Faddeev formalism, the total wavefunction, $\psi^{J}$, can be written as a sum of two-body wavefunctions as follows:
\begin{align}
\psi^{J} = \sum_{i=1}^{3}\psi_i^{J}(\vec{x}_i, \vec{y}_i),
\end{align}
where $\psi^{J}_i(\vec{x}_i, \vec{y}_i)$ are Faddeev amplitudes. 

The two Jacobi coordinates, $\vec{x}_i$ and $\vec{y}_i$, are transformed into a 6-dimensional system of hyperspherical coordinates. These coordinates consist of one hyperradius $\rho$ and 5 angles, which are collectively labelled as $\Omega_5$. In the asymmetric parametrisation, these 5 angular variables are $\Omega_5=(\theta_i, \nu_{x_i}, \nu_{y_i}, \omega_{x_i}, \omega_{y_i})$. The variable $\theta_i \in [0, \pi/2]$ is a hyperangle. The variables $\nu_{x_i} \in [0,\pi]$ and $\omega_{x_i} \in [0, 2 \pi]$ are polar angles related to the Jacobi variable $x_i$, while $\nu_{y_i} \in [0,\pi]$ and $\omega_{y_i} \in [0, 2 \pi]$ are related to $y_i$.

The key to the hyperspherical method are the hyperspherical harmonics, $\mathcal{Y}(\Omega_5)$. These hyperspherical harmonics are constructed as follows:

\begin{align}
\mathcal{Y}_{K_i, m_{x_i}, m_{y_i}}^{\ell_{xi}, \ell_{yi}}(\Omega_5) = \phi_{K_i}^{\ell_{xi}, \ell_{yi}}(\theta_i) Y_{m_{x_i}}^{\ell_{x_i}}(\nu_{x_i},\omega_{x_i}) Y_{m_{y_i}}^{\ell_{y_i}}(\nu_{y_i},\omega_{y_i}),
\end{align}
where $K_i$ are hyperangular momenta, $\phi_{K_i}^{\ell_{xi}, \ell_{yi}}(\theta_i)$ are hyperspherical polynomials, with $Y_{m_{x_i}}^{\ell_{x_i}}(\nu_{x_i},\omega_{x_i})$ and $Y_{m_{y_i}}^{\ell_{y_i}}(\nu_{y_i},\omega_{y_i})$ being spherical harmonics. The orthogonality of the hyperspherical polynomials, $\phi_{K_i}^{\ell_{xi}, \ell_{yi}}(\theta_i)$, renders them a possible basis on which the Faddeev amplitudes can be expanded. After coupling of the angular momenta as outlined in \cite{tho2004}, the Faddeev amplitudes, $\psi_{\alpha_i}^{i,J}(x_i, y_i)$, can be expanded as follows \cite{kie1993, nun1996, tar2004}:
\begin{align}
\label{eq:expansion1}
 \psi_{\alpha_i}^{i,J}(x_i, y_i) =  \sum_{K_i=K_{min}}^{K_{max}} \frac{\chi_{\alpha_i, K_i}^{i,J}(\rho)}{\rho^{5/2}} \phi_{K_i}^{\ell_{xi}, \ell_{yi}}(\theta_i),
\end{align}
where $\alpha_i$ is an abbreviation for the coupling scheme. For a given set of two-body potentials, the Faddeev amplitudes can be fully determined by solving the following system of coupled hyperradial equations for $\chi_{\alpha_i, K_i}^{i,J}(\rho)$ \cite{smi1977, tho2009}:
\begin{align}
\label{eq:hyperradial}
 \left \{ -\frac{\hbar^2}{2m}\frac{d^2}{d \rho^2} + \frac{\hbar^2}{2 m \rho^2}\mathcal{L}_{K_i}(\mathcal{L}_{K_i} +1)  - E  \right \} \chi_{\alpha_i, K_i}^{i,J}(\rho) =- \sum_{ j \alpha_j K_j } V_{\alpha_i K_i, \alpha_j K_j}^{ij} (\rho)       \chi_{\alpha_j, K_j}^{j,J}(\rho),
\end{align}
where $\mathcal{L}_{K_i}=K_i+3/2$ and $V_{\alpha_i K_i, \alpha_j K_j}^{ij} (\rho)$ are couplings. On a basis of normalised associated Laguerre polynomial $\{ R_n(\rho)\}$, the wavefunctions $\chi_{\alpha_i, K_i}^{i,J}(\rho)$ are expanded as follows:
\begin{align} 
\label{eq:expansion2}
\chi_{\alpha_i, K_i}^{i,J}(\rho) = \sum_{n=0}^{N_b} a_{K_i \alpha_i}^{in,J} R_n(\rho),
\end{align} 
where $N_b$ is the size of the model space. Using the two-body potentials presented in the following section, Equation \eqref{eq:hyperradial} is solved with the computer code presented in \cite{tho2004}.

\section{Two-body potentials}
\label{sec:potentials}

The lambda hypertriton is treated as a $p+n+\Lambda$ three-body system. There are therefore three distinct subsystems: the $\Lambda+n$, $\Lambda+p$ and $n+p$ subsystems. At the introduction, it was stated that the aim of this paper is to introduce lambda-proton and lambda-neutron potentials, developed through Gel'fand-Levitan-Marchenko theory, into few-body hypernuclear physics. These potentials were restored through the application of inverse scattering theory on sub-threshold theoretical scattering phases \cite{meo2019}. In order to render these potentials easy to use in this paper and elsewhere, the data representing the effective potentials from \cite{meo2019} were fitted with a sum of three Gaussians, as shown in Equation \eqref{eq:three_gaussians}. 
\begin{align}
\label{eq:three_gaussians}
V_{\Lambda N}(r) = \sum_{i=1}^{3} V_{i} \exp \left \{ \frac{-(r - \mu_i)^2}{\sigma^2_i} \right \}.
\end{align}

As discussed in \cite{meo2019}, this lambda-nucleon potential has 1/4 contribution from the $^1S_0$ channel and 3/4 from the $^3S_1$ channel. Such a superposition, based on Effective Range Theory \cite{bet1949,hac2006}, is applicable to the lambda-nucleon force within other theories, for example in \cite{ans1986}. Spin averaging, based on the strengths of singlet and triplet channel contributions, has a very long history in hypernuclear physics \cite{smi1964,diet1964}. From our application of Gel'fand-Levitan-Marchenko theory, the lambda-proton potential was seen to be stronger than the lambda-neutron potential in both the $^1S_0$ and $^3S_1$ channels, as expected in a lambda-nucleon force. Furthermore, in the effective lambda-nucleon potential given by Equation \eqref{eq:three_gaussians}, the lambda-proton potential is also observed to be stronger than the lambda-nucleon potential, thereby lending credibility to the spin-averaging scheme applied in \cite{meo2019}. It is important to note that these lambda-nucleon potentials do not include lambda-sigma conversion. By clearly stating this weakness, we are providing the user a proper context within which to interpret our results. Some of the widely used potentials from Meson-Exchange theory and Chiral Effective Field theory have undergone many years of improvement to address their own weaknesses. Therefore, the quest on how to handle lambda-sigma conversion within inverse scattering theory remains an open problem. The goal of these series of studies, as stated in \cite{meo2019}, is to present inverse scattering theory as a \textit{complement} to Meson-Exchange theory and Chiral Effective Field theory.

\begin{figure}[h!]
\centering
\subfigure[]{\includegraphics[width=0.7\linewidth]{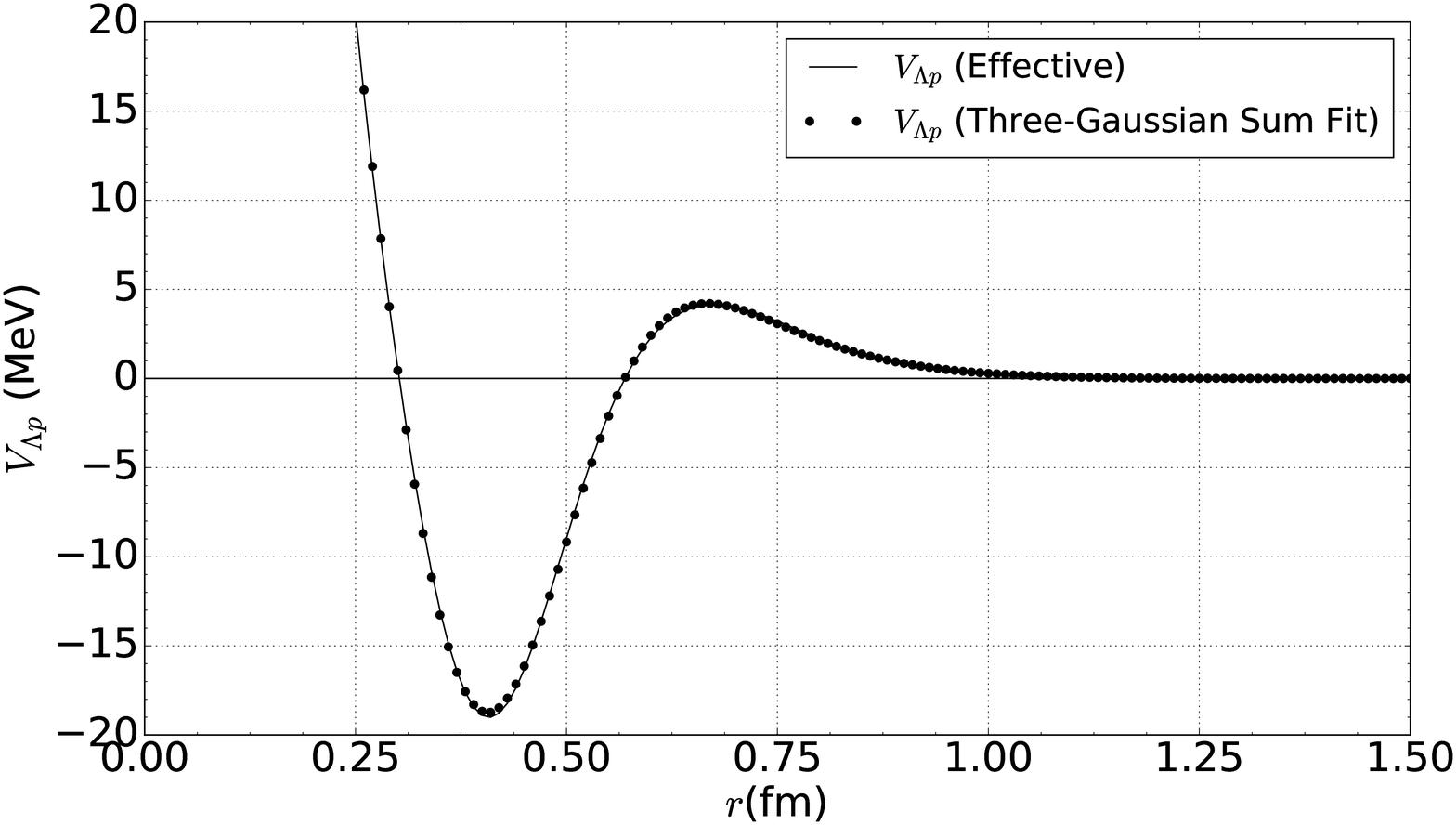}}
\subfigure[]{\includegraphics[width=0.7\linewidth]{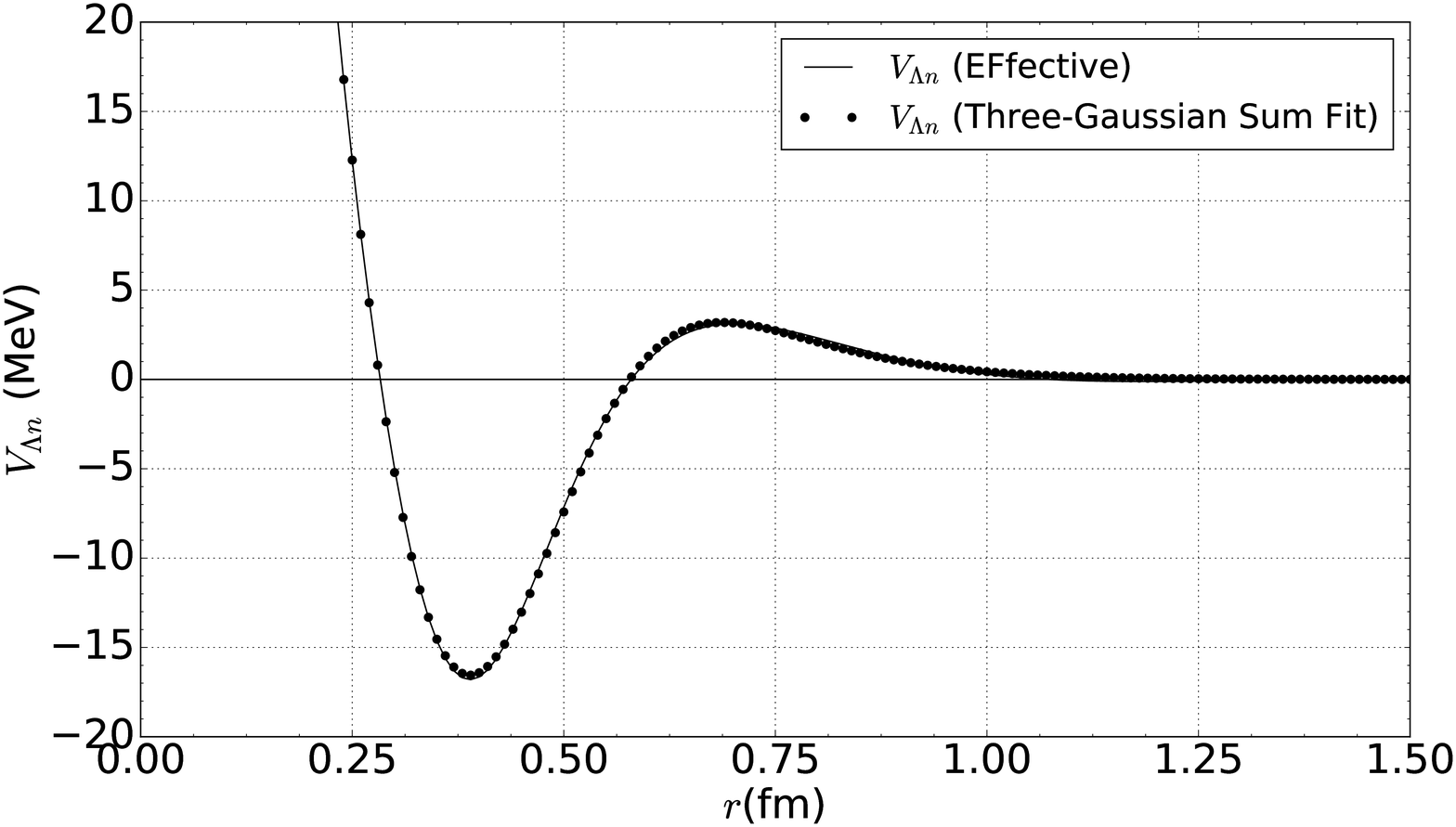}}
\caption{\label{fig:fitted} Comparison of the three-term Gaussian fits to the data from \cite{meo2019}.}
\end{figure}

The parameters $V_{i}, \mu_i$ and $\sigma_i$ in Equation \eqref{eq:three_gaussians} were determined through a Nonlinear Least Squares Fit. Minimization of the objective functional was carried out using the Levenberg-Marquardt algorithm. After convergence, the estimated parameters that were obtained are displayed in Table \ref{fit}. The error (uncertainty) in the estimates of $\mu_i$ and $\sigma_i$ are indicated. A comparison of these three-term Gaussian fits with the data from \cite{meo2019} is shown in Figure \ref{fig:fitted}. For ease of reference, this version of lambda-nucleon potentials from inverse scattering theory shall be referred to as GLM-YN0 potentials.

\begin{table}[h!]
\centering
\caption[]{Estimates of fit parameters of $\Lambda$-proton  and $\Lambda$-neutron effective potentials, $V_{\Lambda p}$ and $V_{\Lambda n}$, respectively. The error in $\mu_i$ and $\sigma_i$ estimates are indicated.}
\begin{tabular}{p{1.0cm}p{3.3cm}p{3.3cm}p{3.4cm}}
           &                  &   $\Lambda p$             &                               \\  \hline
           &  $V_i$ /MeV      &   $\mu_i$/ fm             &  $\sigma_i$/fm                \\ \hline \hline
$i=1$      & $45.88      $    &   $0.1148 \pm 0.0006601$  &  $-0.3932 \pm 0.0008502$      \\ 
$i=2$      & $8.106e+07  $    &   $-1.193 \pm 0.001948$   &  $0.3575 \pm 0.0005306$       \\ 
$i=3$      & $-47.04     $    &   $0.3748 \pm 0.0001386$  &  $0.1667 \pm 0.0002179$       \\   \hline
           &                  &                           &                               \\
           &                  &                           &                               \\
           &                  &    $\Lambda n$            &                               \\ \hline
           &  $V_i$ /MeV      &   $\mu_i$/ fm             &  $\sigma_i$/fm                \\ \hline \hline 
$i=1$      & $186.9      $    &   $- 0.3476 \pm 0.001364$ &  $-0.5469 \pm 0.001125$       \\ 
$i=2$      & $6.74e+04   $    &   $- 0.383 \pm 0.001433$  &  $0.191 \pm 0.0005638$        \\ 
$i=3$      & $- 52.14    $    &   $0.3243 \pm 0.0001977$  &  $0.2013 \pm 0.0002466$       \\   \hline
\end{tabular}
\label{fit}
\end{table}

For the neutron-proton subsystem, the spin-averaged Malfliet-Tjon potential (MT-V) was used \cite{mal1969}. The MT-V potential is a sum of two Yukawa functions, as shown in Equation \eqref{eq:mtv}. The parameters used for this spin-independent potential, displayed in Table \ref{tab:mtv}, are from \cite{zab1982}. This simple nucleon-nucleon potential, with only a central term and no tensor or spin-orbit or momentum-dependent terms, has traditionally been used as a tool in the benchmarking or comparison of various few-body calculations in nuclear physics, even in recent works \cite{myo2017, gar2020}. 
\begin{align}
\label{eq:mtv}
V_{np}(r)= \sum_{i=1}^{2} \frac{V_{i}}{r} \exp (-\beta_i r). 
\end{align}

\begin{table}[h!]
\centering
\caption[Parameters for spin-averaged Malfliet-Tjon-V potential, MT-V ($V_{np}$)]{Parameters for spin-averaged Malfliet-Tjon potential, MT-V ($V_{np}$). These parameters are from Zabolitzky \cite{zab1982}.}
\begin{tabular}{p{3cm}p{3cm}p{3cm}}\hline
           &  $V_i/\text{MeV}\cdot \text{fm} $ &   $\beta_i$/fm$^{-1}$             \\  \hline 
$i=1$      &  1458.05                          &   3.11                           \\ 
$i=2$      &  -578.09                          &   1.55                           \\   \hline
\end{tabular}
\label{tab:mtv}
\end{table}

\section{Results and discussion}

The results of three-body Faddeev calculations for the ground state of the hypertriton ($J=1/2^{+}$), using the potentials in Section \ref{sec:potentials}, are reported here. The masses used for the proton and neutron are $m_p=1.007276466$u and $m_n=1.008664915$u, respectively \cite{moh2016}. The mass of the lambda hyperon is calculated from its energy equivalence i.e. $m_{\Lambda}=1115.683 / 931.5 = 1.198$u. These masses enter the computation through the Jacobi coordinates (Equation \eqref{eq:jacobi_coord}), which are transformed into hyperspherical variables. The mass parameter, $\hbar^2/ 2m $, in Equation \eqref{eq:hyperradial} was computed as follows:
\begin{align}
\frac{\hbar^2}{2m} &= \frac{(\hbar c)^2}{2m c^2}, \notag \\
                  &= \frac{(197.3 \,\, \text{MeV}\cdot\text{fm})^2}{2(939.0 \,\, \text{MeV})} =  20.7281 \,\, \text{MeV}\cdot\text{fm}^2, \notag
\end{align}
where $mc^2 = 939.0$ MeV is the energy equivalence of the nucleon mass. The maximum values of the quantum numbers ($K_i, S_{xi}, \ell_{xi}, \ell_{yi}$) used in defining the channels are $K_{max}=8, S_{xmax}=1.0, \ell_{xmax}=2$ and $\ell_{ymax}=2$.  

\begin{figure}[h!]
\centering
\subfigure[]{\includegraphics[width=0.7\linewidth]{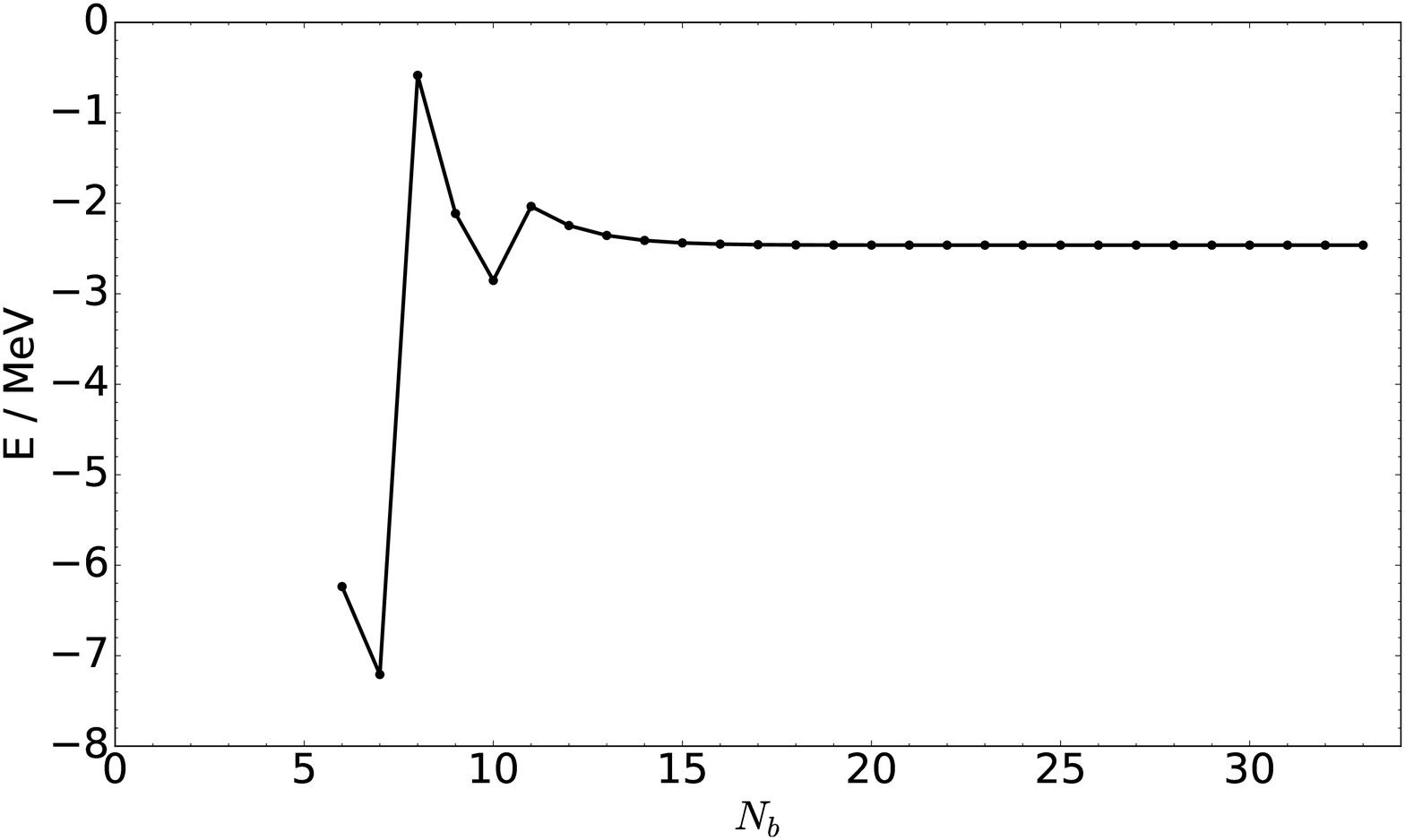}}
\subfigure[]{\includegraphics[width=0.7\linewidth]{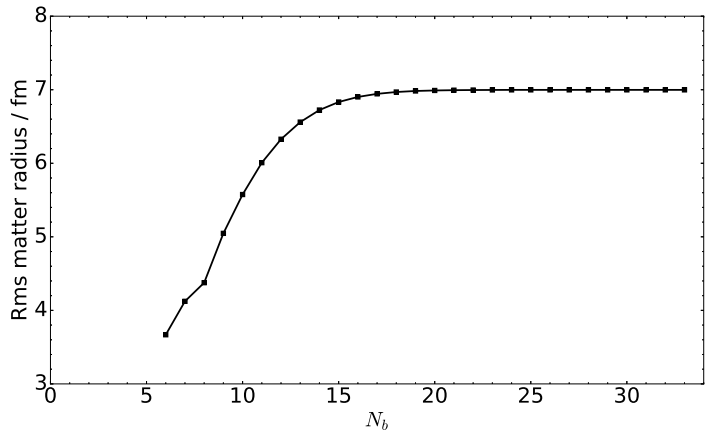}}
\caption{\label{fig:hypertriton_convergence} Convergence of hypertriton ground state binding energy ($E$) and root-mean-square radius with size of model space ($N_b$)}
\end{figure}

The three-body computations, done through an inverse iteration, found a $J=1/2^{+}$ bound state of $-2.462$ MeV and a r.m.s. matter radius of $7.00$ fm for the hypertriton. Table \ref{tab:hypertriton_convergence} and Figure \ref{fig:hypertriton_convergence}  show the convergence behaviour of the binding energy and r.m.s. radius. The convergence behaviour shown in these results is identical to that in \cite{nan2018}, where the Non-Symmetrized Hyperspherical Harmonics method was used in computing the binding energy of a triton. A comparison of the results presented in this paper with those from experimental studies and other theoretical predictions is shown in Table \ref{tab:hypertriton_compare}. The r.m.s. radius reported in this paper is much larger than the value of 4.9 fm presented in \cite{nem2000}. This larger r.m.s. radius is a result of the underbinding of the deuteron by the Malfliet-Tjon-V potential (MT-V) \cite{mal1969}.

\begin{table}[h!]
\centering
\caption[Convergence of hypertriton ground state binding energy and root-mean-square matter radius with size of model space.]{Convergence of hypertriton ground state binding energy and root-mean-square matter radius with size of model space.}
\begin{tabular}{p{3cm} p{3cm} p{5cm}}\hline 
    $N_b$                  & $E$ / MeV           & R.m.s matter radius / fm    \\ \hline \hline
06                         & -6.235987           & 3.668                             \\
08                         & -0.585377           & 4.376                             \\
10                         & -2.852280           & 5.575                             \\
12                         & -2.245663           & 6.328                             \\
14                         & -2.410353           & 6.724                              \\
16                         & -2.450869           & 6.903                                \\
18                         & -2.459966           & 6.969                                \\
20                         & -2.461853           &  6.990    \\
22                         & \textbf{-2.462}224           &  6.996    \\
24                         & \textbf{-2.462}294           &  6.998    \\
26                         & \textbf{-2.462}307           &  6.998    \\
28                         & \textbf{-2.462}309           &  6.998    \\
30                         & \textbf{-2.462}309           &  6.998    \\
32                         & \textbf{-2.462}310           &  6.998   \\ \hline
\end{tabular}
\label{tab:hypertriton_convergence}
\end{table}

\begin{table}[h!]
\centering
\caption[Hypertriton binding energy from our three-body calculation, compared with results from other three-body studies and from experiments.]{Hypertriton binding energy from our three-body calculation, compared with results from other three-body studies and from experiments. The $\Lambda$-nucleon potentials used are indicated in parentheses.}
\begin{tabular}{p{7.7cm} p{2.5cm} }\hline 
                                                                & E / MeV                \\ \hline \hline
  Experiment 1 \cite{jur1973,dav2005} (Emulsion)                & $-2.35\pm 0.05$              \\ 
  Experiment 2 \cite{key1970} (Helium bubble chambers)          & $-2.47\pm 0.31$              \\ 
  \textbf{This paper (GLM-YN0)}                                 & $\textbf{-2.462}$          \\ 
  Fujiwara \textit{et al.} \cite{fuj2004b} (FSS)          & $-3.134$                     \\ 
  Fujiwara \textit{et al.} \cite{fuj2004b} (fss2)         & $-2.514$                         \\ 
  Fujiwara \textit{et al.} \cite{fuj2008} (fss2, modified) & $-2.487$                     \\ 

  Ferrari \textit{et al.} \cite{fer2017} (NSC97f)            & $-2.41(2)$             \\ 

  Tominaga \& Ueda         \cite{tom2001,tom1998} (Ehime 00A, single)& $-2.35$             \\ 
  Miyagawa \textit{et al.} \cite{miy1995} (NSC97f)        & $-2.37$                      \\ 

  Polinder \textit{et al.}  \cite{pol2006, hai2013b} ($\chi$EFT LO)& $-2.34 - -2.36$                       \\ 
  Haidenbauer \cite{hai2012, hai2013b} ($\chi$EFT NLO)             & $-2.31 - -2.34$                      \\ 
  Haidenbauer \cite{hai2013b} (NSC97f)                    & $-2.30$                           \\ 
  Haidenbauer \cite{hai2012, hai2013b} (J\"{u}lich '04)      & $-2.27$                            \\ 
  Miyagawa \textit{et al.} \cite{rij1999, miy2000a} (NSC97a-d) & Unbounded                 \\ 
  Miyagawa \& Gl\"ockle    \cite{miy1993a} (J\"{u}lich A)      & Unbounded               \\ \hline
\end{tabular}
\label{tab:hypertriton_compare}
\end{table}

\begin{figure}[h!]
\centering
\subfigure[]{\includegraphics[width=0.45\linewidth]{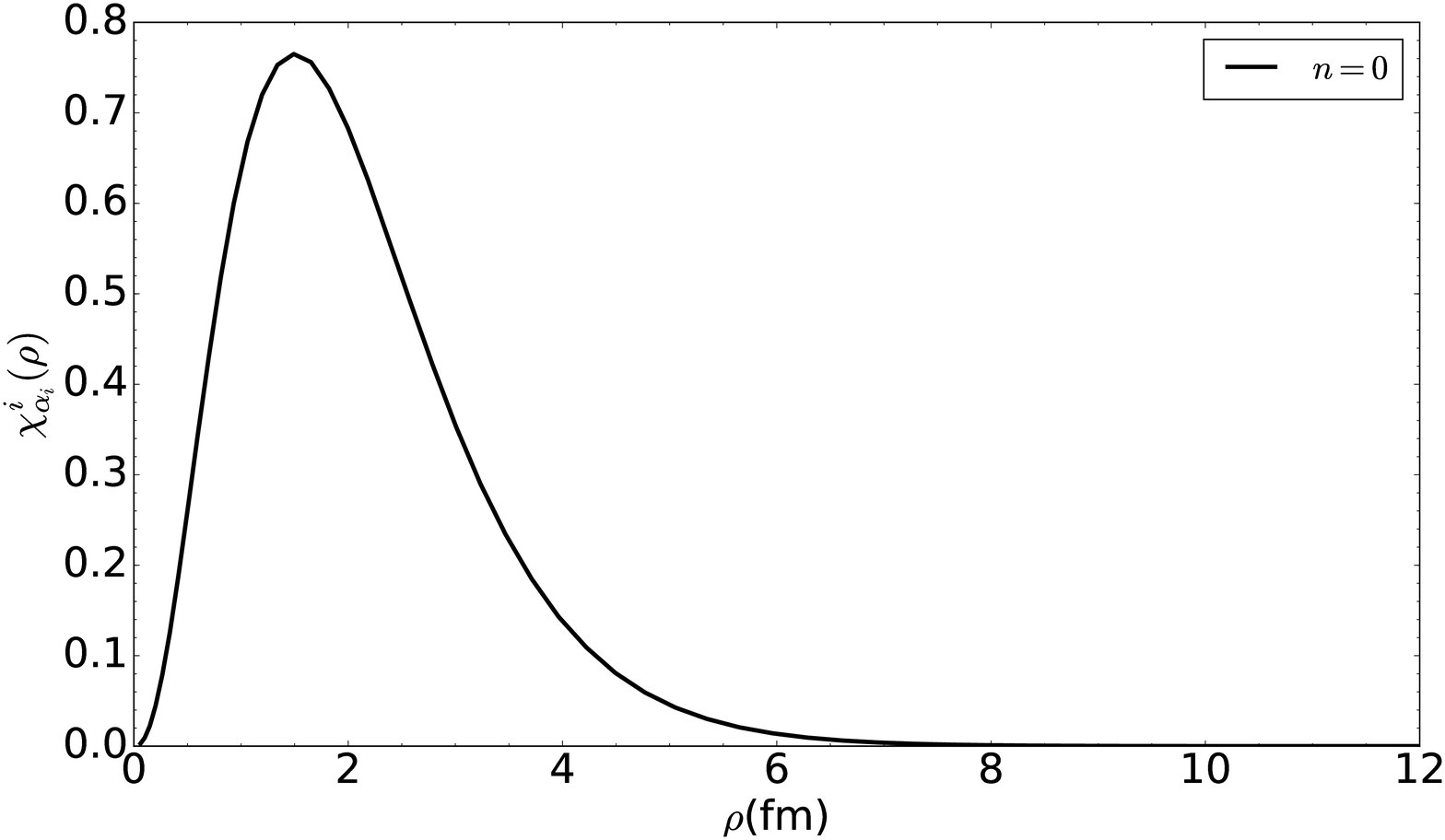}}
\subfigure[]{\includegraphics[width=0.45\linewidth]{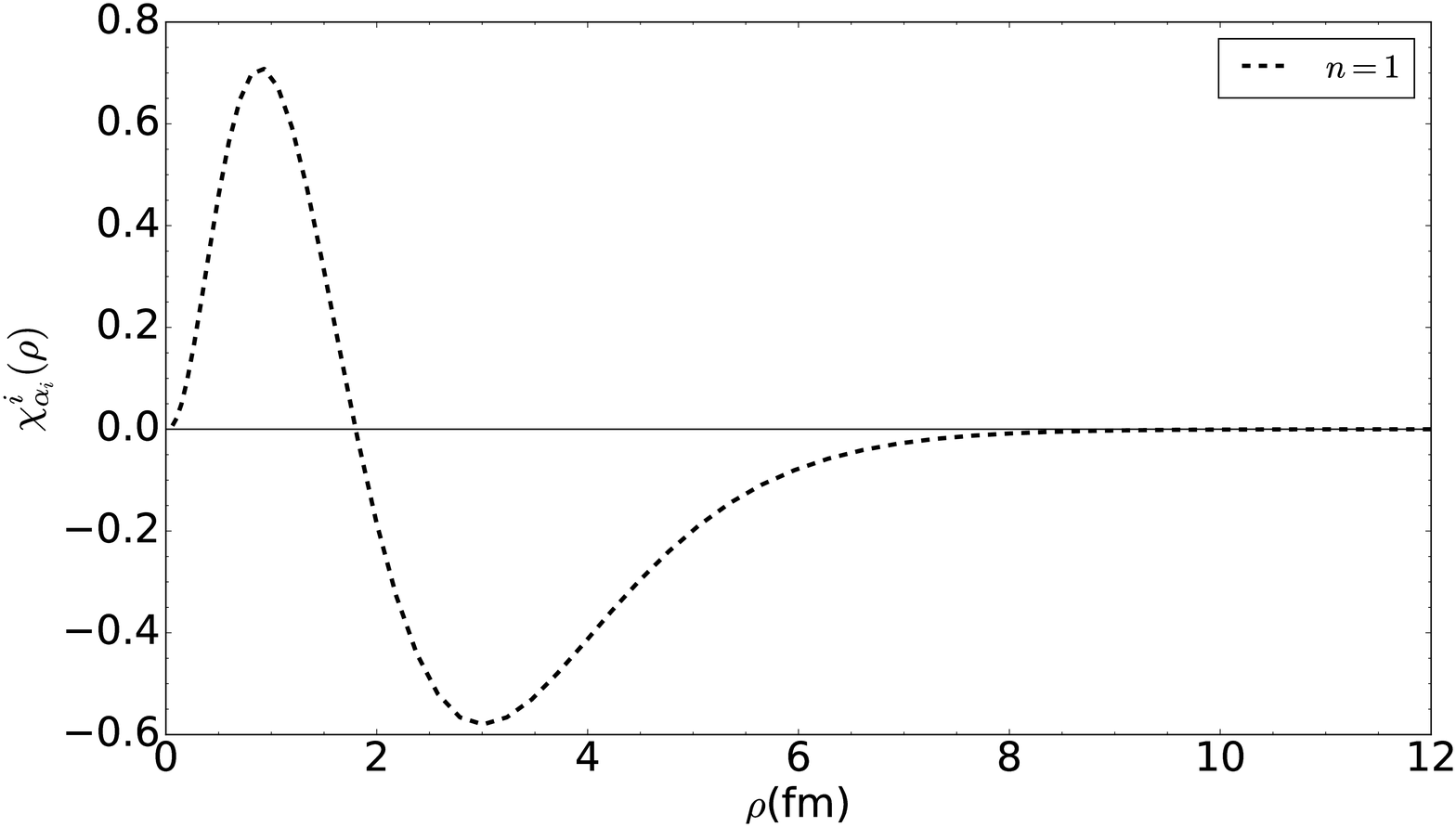}}

\subfigure[]{\includegraphics[width=0.45\linewidth]{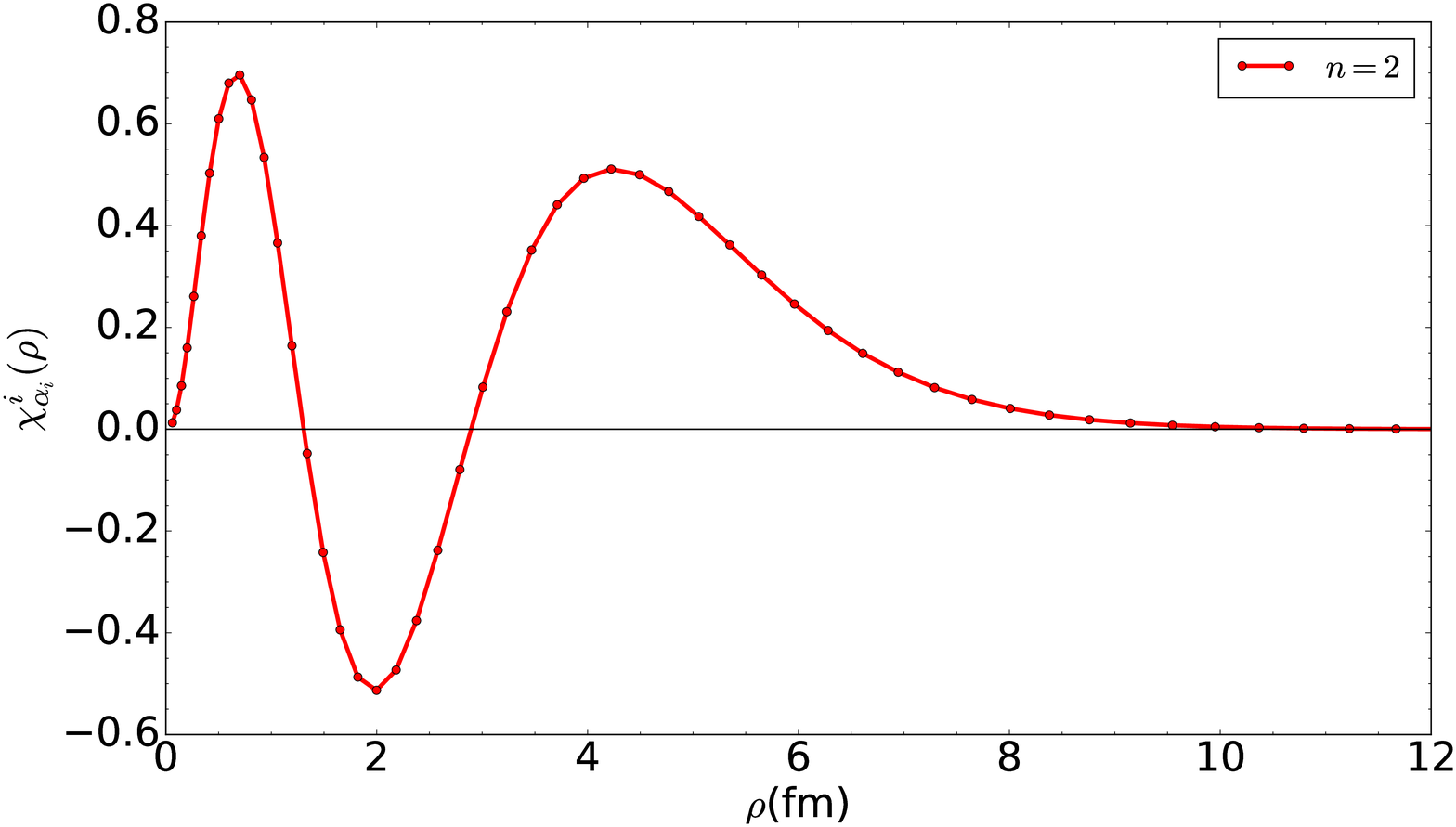}}
\subfigure[]{\includegraphics[width=0.45\linewidth]{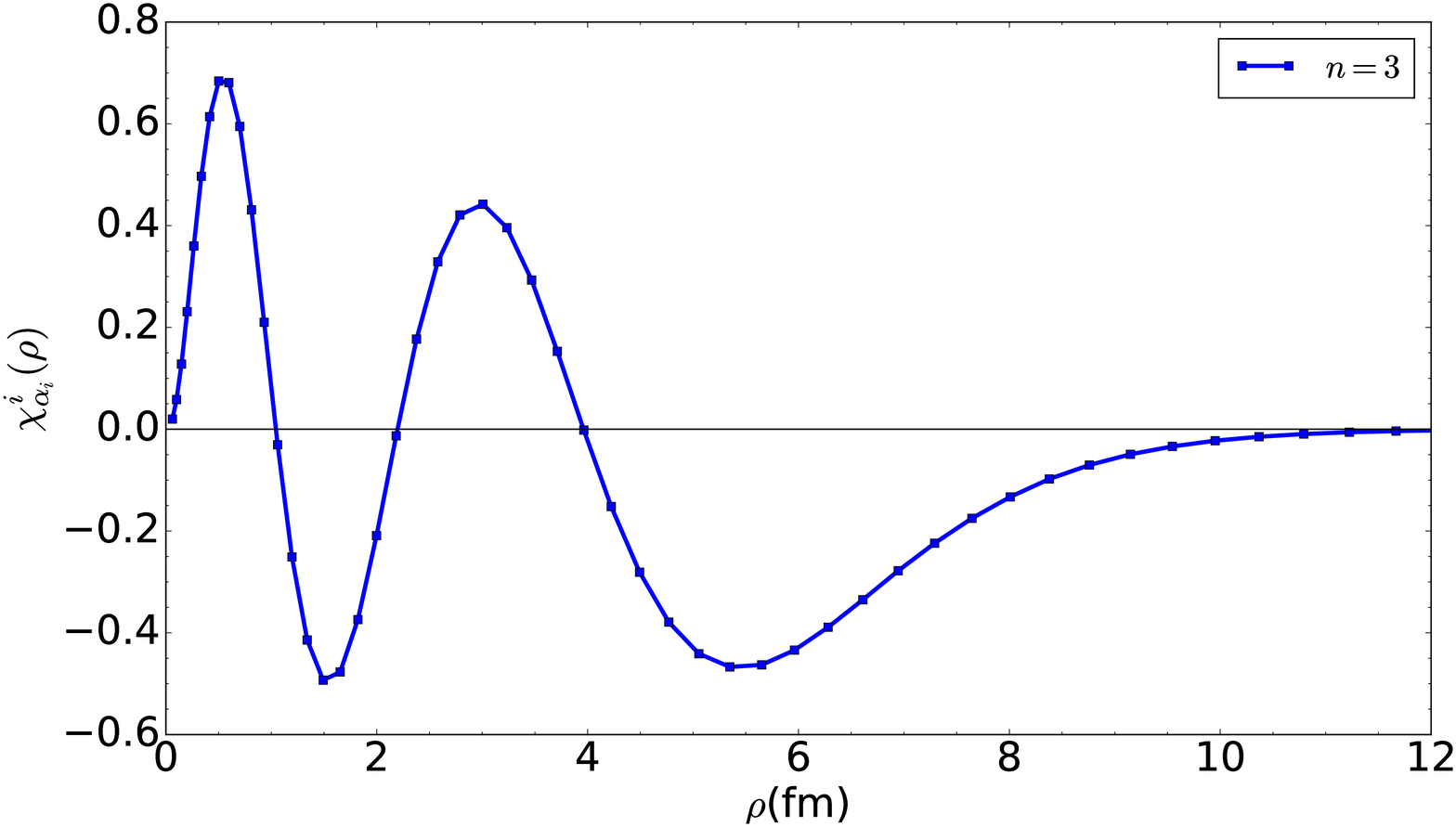}}
\caption{\label{fig:radial_wavefunctions} First four hyperradial wavefunctions in the dominant channel. These wavefunctions become more oscillatory as one progresses through the terms in the expansion in Equation \eqref{eq:expansion2}.}
\end{figure}

The hyperradial behaviour is obtained from the contribution of all channels, identified by the quantum numbers $\alpha=\{ K, L, S_{x}, l_{x}, \ell_{y}\}$, in that order. For the channel with the dominant contribution, the first four hyperradial wavefunctions are shown in Figure \ref{fig:radial_wavefunctions}. As one progresses within this channel, these hyperradial wavefunctions are observed to become increasingly oscillatory. 

\section{Conclusions}

The ground state binding energy and root-mean-square radius of the lambda hypertriton were computed through the Differential Faddeev Equations in hyperspherical variables. The lambda hypertriton was treated as a proton+neutron+lambda three-body system. In the lambda-proton and lambda-neutron subsystems, the potentials used (GLM-YN0 potentials) have their roots in Gel'fand-Levitan-Marchenko theory while the neutron-proton potential is the simple Malfliet-Tjon-V potential. The results obtained are -2.462 MeV and 7.00 fm for the binding energy and root-mean-square radius, respectively. The convergence of the few-body calculations using these GLM-YN0 lambda-nucleon potentials was also observed to be satisfactory. In order to assist in the proper interpretation of these results, it is important to specify that the GLM-YN0 potentials do not have a lambda-sigma conversion component. Due to the absence of lambda-sigma conversion, the potentials \textit{may} not be suitable for heavy hypernuclei. Nonetheless, these computations are significant because they represent the first application of hyperon-nucleon potentials from Gel'fand-Levitan-Marchenko theory in hypernuclear few-body physics. The results show that inverse scattering theory can play a useful role as a complement to Meson-Exchange theory and Chiral Effective Field theory in probing the hyperon-nucleon force. Further computations are required to assess these new lambda-proton and lambda-neutron potentials for conformity with other known features of the lambda-nucleon force, for example, Charge Symmetry Breaking.

\section*{References}

\bibliographystyle{iopart-num} 
\bibliography{nuclearref}

\providecommand{\newblock}{}
\begin{thebibliography}{10}
\expandafter\ifx\csname url\endcsname\relax
  \def\url#1{{\tt #1}}\fi
\expandafter\ifx\csname urlprefix\endcsname\relax\def\urlprefix{URL }\fi
\providecommand{\eprint}[2][]{\url{#2}}

\bibitem{dav2005}
Davis D~H 2005 {\em Nuclear Physics A\/} {\bf 754} 3 -- 13 proceedings of the
  Eighth International Conference on Hypernuclear and Strange Particle Physics

\bibitem{dal2005}
Dalitz R~H 2005 {\em Nuclear Physics A\/} {\bf 754} 14 -- 24 proceedings of the
  Eighth International Conference on Hypernuclear and Strange Particle Physics

\bibitem{deSwart1971}
de~Swart J~J, Nagels M~M, Rijken T~A and Verhoeven P~A 1971 {\em Springer
  Tracts in Modern Physics\/} vol \textbf{60} ed H{\"o}hler G (Berlin,
  Heidelberg: Springer Berlin Heidelberg) pp 138--203

\bibitem{deSwart1996}
de~Swart J~J, Klomp R~A~M~M, Rentmeester M~C~M and Rijken T~A 1996 {\em
  Few-Body Problems in Physics '95: In memoriam Professor Paul Urban\/} ed
  Guardiola R (Vienna: Springer Vienna) pp 438--447

\bibitem{rij1993}
Rijken T~A 1993 {\em Few-Body Problems in Physics '93, Supplementum 7.
  Proceedings of the XIVth European Conference on Few-Body Problems in
  Physics\/} ed Bakker B~L~G and van Dantzig R (Wien: Springer Verlag) pp 1--12

\bibitem{rij2001}
Rijken T~A 2001 {\em Nuclear Physics A\/} {\bf \textbf{691}} 322 -- 328 proc.
  7th Int. Conf. on Hypernuclear and Strange Particle Physics

\bibitem{hol1989}
Holzenkamp B, Holinde K and Speth J 1989 {\em Nuclear Physics A\/} {\bf
  \textbf{500}} 485 -- 528

\bibitem{reu1992}
Reuber A, Holinde K and Speth J 1992 {\em Czechoslovak Journal of Physics\/}
  {\bf \textbf{42}} 1115--1135

\bibitem{hai2005}
Haidenbauer J and Mei\ss{}ner U~G 2005 {\em Physical Review C\/} {\bf
  \textbf{72}} 044005

\bibitem{fuj1996a}
Fujiwara Y, Nakamoto C and Suzuki Y 1996 {\em Physical Review C\/} {\bf
  \textbf{54}}(5) 2180--2200

\bibitem{fuj1996b}
Fujiwara Y, Nakamoto C and Suzuki Y 1996 {\em Physical Review Letters\/} {\bf
  \textbf{76}}(13) 2242--2245

\bibitem{fuj2001}
Fujiwara Y, Kohno M, Nakamoto C and Suzuki Y 2001 {\em Phys. Rev. C\/} {\bf
  64}(5) 054001

\bibitem{pol2006}
Polinder H, Haidenbauer J and Mei{\ss}ner U~G 2006 {\em Nuclear Physics A\/}
  {\bf 779} 244 -- 266

\bibitem{pol2007}
Polinder H, Haidenbauer J and Mei{\ss}ner U~G 2007 {\em Physics Letters B\/}
  {\bf 653} 29 -- 37

\bibitem{hai2013}
Haidenbauer J, Petschauer S, Kaiser N, Mei{\ss}ner U~G, Nogga A and Weise W
  2013 {\em Nuclear Physics A\/} {\bf 915} 24 -- 58

\bibitem{ada2018}
Adamczyk L {\em et~al.\/} 2018 {\em Phys. Rev. C\/} {\bf 97}(5) 054909 (STAR
  Collaboration)

\bibitem{gal2018b}
Gal A and Garcilazo H 2019 {\em Physics Letters B\/} {\bf 791} 48 -- 53

\bibitem{tro2018}
Trogolo S 2019 {\em Nuclear Physics A\/} {\bf 982} 815 -- 818 the 27th
  International Conference on Ultrarelativistic Nucleus-Nucleus Collisions:
  Quark Matter 2018

\bibitem{xu2017}
Xu Y 2017 {\em JPS Conf. Proc.\/} {\bf 17} 021005

\bibitem{meo2019}
Meoto E~F and Lekala M~L 2019 {\em Journal of Physics Communications\/} {\bf 3}
  095018

\bibitem{tho2004}
Thompson I~J, Nunes F~M and Danilin B~V 2004 {\em Computer Physics
  Communications\/} {\bf 161} 87 -- 107

\bibitem{kie1993}
Kievsky A, Viviani M and Rosati S 1993 {\em Nuclear Physics A\/} {\bf 551} 241
  -- 254

\bibitem{nun1996}
Nunes F, Christley J, Thompson I, Johnson R and Efros V 1996 {\em Nuclear
  Physics A\/} {\bf 609} 43 -- 73

\bibitem{tar2004}
Tarutina T, Thompson I and Tostevin J 2004 {\em Nuclear Physics A\/} {\bf 733}
  53 -- 66

\bibitem{smi1977}
Smirnov Y~F and Shitikova K~V 1977 {\em Sov. J. Particles Nucl.\/} {\bf 8}

\bibitem{tho2009}
Thompson I~J and Nunes F~M 2009 {\em Nuclear Reactions for Astrophysics:
  Principles, Calculation and Applications of Low-Energy Reactions\/}
  (Cambridge: Cambridge University Press)

\bibitem{bet1949}
Bethe H~A 1949 {\em Phys. Rev.\/} {\bf 76}(1) 38--50

\bibitem{hac2006}
Hackenburg R~W 2006 {\em Phys. Rev. C\/} {\bf 73}(4) 044002

\bibitem{ans1986}
Ansari H~H, Shoeb M and Khan M~Z~R 1986 {\em Journal of Physics G: Nuclear
  Physics\/} {\bf 12} 1369

\bibitem{smi1964}
Smith D~R and Downs B~W 1964 {\em Phys. Rev.\/} {\bf 133}(2B) B461--B465

\bibitem{diet1964}
Dietrich K, Mang H and Folk R 1964 {\em Nuclear Physics\/} {\bf 50} 177 -- 201

\bibitem{mal1969}
Malfliet R~A and Tjon J~A 1969 {\em Nuclear Physics A\/} {\bf 127} 161 -- 168

\bibitem{zab1982}
Zabolitzky J~G, Schmidt K~E and Kalos M~H 1982 {\em Phys. Rev. C\/} {\bf 25}(2)
  1111--1113

\bibitem{myo2017}
Myo T, Toki H, Ikeda K, Horiuchi H and Suhara T 2017 {\em Phys. Rev. C\/} {\bf
  95}(4) 044314

\bibitem{gar2020}
Garcilazo H, Valcarce A and Vijande J 2020 {\em Chinese Physics C\/} {\bf 44}
  024102

\bibitem{moh2016}
Mohr P~J, Newell D~B and Taylor B~N 2016 {\em Rev. Mod. Phys.\/} {\bf 88}(3)
  035009

\bibitem{nan2018}
Nannini A and Marcucci L~E 2018 {\em Frontiers in Physics\/} {\bf 6} 122

\bibitem{nem2000}
Nemura H, Suzuki Y, Fujiwara Y and Nakamoto C 2000 {\em Progress of Theoretical
  Physics\/} {\bf 103} 929--958

\bibitem{jur1973}
Juri\v{c} M {\em et~al.\/} 1973 {\em Nuclear Physics B\/} {\bf 52} 1 -- 30

\bibitem{key1970}
Keyes G, Derrick M, Fields T, Hyman L~G, Fetkovich J~G, McKenzie J, Riley B and
  Wang I~T 1970 {\em Phys. Rev. D\/} {\bf 1}(1) 66--77

\bibitem{fuj2004b}
Fujiwara Y, Miyagawa K, Kohno M and Suzuki Y 2004 {\em Nuclear Physics A\/}
  {\bf 738} 382 -- 386 proceedings of the 8th International Conference on
  Clustering Aspects of Nuclear Structure and Dynamics

\bibitem{fuj2008}
Fujiwara Y, Suzuki Y, Kohno M and Miyagawa K 2008 {\em Phys. Rev. C\/} {\bf
  77}(2) 027001

\bibitem{fer2017}
Ferrari~Ruffino F, Lonardoni D, Barnea N, Deflorian S, Leidemann W, Orlandini G
  and Pederiva F 2017 {\em Few-Body Systems\/} {\bf 58} 113

\bibitem{tom2001}
Tominaga K and Ueda T 2001 {\em Nuclear Physics A\/} {\bf 693} 731 -- 754

\bibitem{tom1998}
Tominaga K, Ueda T, Yamaguchi M, Kijima N, Okamoto D, Miyagawa K and Yamada T
  1998 {\em Nuclear Physics A\/} {\bf 642} 483 -- 505

\bibitem{miy1995}
Miyagawa K, Kamada H, Gl\"ockle W and Stoks V 1995 {\em Phys. Rev. C\/} {\bf
  51}(6) 2905--2913

\bibitem{hai2013b}
Haidenbauer J 2013 {\em Nuclear Physics A\/} {\bf 914} 220 -- 230 xI
  International Conference on Hypernuclear and Strange Particle Physics
  (HYP2012)

\bibitem{hai2012}
Haidenbauer J 2012 {\em Proceedings of science\/} The 7th International
  Workshop on Chiral Dynamics, August 6 -10, 2012

\bibitem{rij1999}
Rijken T~A, Stoks V~G~J and Yamamoto Y 1999 {\em Physical Review C\/} {\bf
  \textbf{59}}(1) 21--40

\bibitem{miy2000a}
Miyagawa K, Kamada H, Gloeckle W, Yamamura H, Mart T and Bennhold C 2000 {\em
  Few-Body Systems Suppl.\/} {\bf 12} 324

\bibitem{miy1993a}
Miyagawa K and Gl\"ockle W 1993 {\em Phys. Rev. C\/} {\bf 48}(6) 2576--2584

\end{thebibliography}
\end{document}